\documentclass[aps,twocolumn,prb]{revtex4}
\usepackage{amssymb}
\usepackage{amsmath}
\usepackage[dvips]{graphicx}

\begin{document}

\title{Reversal modes in magnetic nanotubes}
\author{P. Landeros, S. Allende, J. Escrig, E. Salcedo, and D. Altbir}
\affiliation{Departamento de F\'{\i}sica, Universidad de Santiago de Chile, USACH, Av.
Ecuador 3493, Santiago, Chile}
\author{ E. E. Vogel}
\affiliation{Departamento de Ciencias F\'{\i}sicas, Universidad de la Frontera, Casilla
54-D, Temuco, Chile}
\keywords{magnetic nanotubes, ferromagnetic tubes, hollow nanowires,
nanomagnetism, magnetic switching, vortex wall, transverse wall}
\pacs{75.75.+a,75.10.-b,75.60.Jk}

\begin{abstract}
The magnetic switching of ferromagnetic nanotubes is investigated as a
function of their geometry. Two independent methods are used: Numerical
simulations and analytical calculations. It is found that for long tubes the
reversal of magnetization is achieved by two mechanism: The propagation of a
transverse or a vortex domain wall depending on the internal and external
radii of the tube.
\end{abstract}

\maketitle

During the last decade, interesting properties of magnetic nanowires have
attracted great attention. Besides the interest in their basic properties,
there is evidence that they can be used in the production of new devices.\
More recently magnetic nanotubes have been grown\cite%
{tubes1,tubes2,Daub,tubes3} motivating a new research field. Magnetic
measurements,\cite{Daub} numerical simulations\cite{tubes3} and analytical
calculations\cite{Escrigt} on such tubes have identified two main states: an
in-plane magnetic ordering, namely the flux-closure vortex state, and a
uniform state with all the magnetic moments pointing parallel to the axis of
the tube. An important problem is to establish the way and conditions for
reversing the orientation of the magnetization. Although the reversal
process is well known for ferromagnetic nanowires,\cite{wires1, wires2,
wires3, wires4,NielschIEEE} the equivalent phenomenon in nanotubes has been
poorly explored so far in spite of some potential advantages over solid
cylinders. Nanotubes exhibit a core-free magnetic configuration leading to
uniform switching fields, guaranteeing reproducibility,\cite{tubes3,Escrigt}
and due to their low density they can float in solutions making them
suitable for applications in biotechnology (see [1] and refs. therein).

Let us consider a ferromagnetic nanotube in a state with the magnetization $%
\mathbf{M}$ along the tube axis. A constant and uniform magnetic field is
then imposed antiparallel to $\mathbf{M}$. After some delay time the
magnetization reversal (MR) will start at any end. MR or magnetic switching
can occur by means of different mechanisms, depending on the geometrical
parameters of the tube. In this paper we will focus on the reversal process
by means of two different but complementary approaches: numerical
simulations and analytical calculations. Their mutual agreement sustains the
results reported in this study.\bigskip

\textbf{Numerical Simulations}. Geometrically, tubes are characterized by
their external and internal radii, $R$ and $a$ respectively, and height, $H$%
. It is convenient to define the ratio $\beta \equiv a/R$, so that $\beta =0$
represents a solid cylinder and $\beta \rightarrow 1$ correspond to a very
narrow tube. The internal energy, $E$, of a nanotube with $N$ magnetic
moments can be written as
\begin{equation*}
E=\sum_{i=1}^{N}\sum_{j>i}^{N}\left( E_{ij}-J_{ij}\boldsymbol{\hat{\mu}}%
_{i}\cdot \boldsymbol{\hat{\mu}}_{j}\right) +E_{a}\,,
\end{equation*}%
\noindent where $E_{ij}$ is the dipolar energy given by $E_{ij}=\left[
\boldsymbol{\mu }_{i}\cdot \boldsymbol{\mu }_{j}-3(\boldsymbol{\mu }%
_{i}\cdot \boldsymbol{\hat{n}}_{ij})(\boldsymbol{\mu }_{j}\cdot \boldsymbol{%
\hat{n}}_{ij})\right] /r_{ij}^{3}\ $, with $r_{ij}$ the distance between the
magnetic moments $\boldsymbol{\mu }$\textbf{$_{i}$} and $\boldsymbol{\mu }%
_{j}$, $\boldsymbol{\hat{\mu}}_{i}$ the unit vector along the direction of $%
\boldsymbol{\mu }_{i}$ and $\boldsymbol{\hat{n}}_{ij}$ the unit vector along
the direction that connects $\boldsymbol{\mu }_{i}$ and $\boldsymbol{\mu }%
_{j}$. $J_{ij}=J$ is the exchange coupling constant between nearest
neighbors and $J_{ij}=0$ otherwise. $E_{a}=-\sum_{i=1}^{N}\boldsymbol{\mu }%
_{i}\cdot \boldsymbol{H}_{a}$ is the contribution of the external
magnetic field. In this paper we are interested in soft magnetic
materials, in which case anisotropy can be safely neglected.\cite%
{tubes1,tubes3} In our simulations we consider Nickel tubes with $|%
\boldsymbol{\mu }_{i}|=0.61$ $\mu _{B}$, lattice parameter $a_{0}=3.52$ \AA\ %
and $J=3.5$ meV. Dimensions are $R=15$ nm, length $H=0.5$ $\mu $m, assuming
a growth along the [100] direction of a fcc lattice. Tubes in the above
mentioned range of sizes have at least $10^{8}$ atoms, and then numerical
simulations at the atomic level are out of reach with present computational
facilities. In order to reduce the number of interacting atoms, we make use
of the scaling technique presented before,\cite{scaling1} applied to the
calculation of the phase diagram of cylindrical particles. Authors show that
such diagram is equivalent to the one of a smaller particle with linear
dimensions $d^{\prime }=d\chi ^{\eta }$ being $\chi <1$ and $\eta \approx
0.55$, if the exchange constant has been also scaled as $J^{\prime }=\chi J$%
. It has also been shown\cite{fastmontecarlo} that the scaling relations can
be used together with Monte Carlo (MC) simulations to obtain a general
magnetic state of a nanoparticle. We use this idea starting from the desired
value for the total number of interacting particles we can deal with, which
based on the computational facilities currently available, is around $3000$.
With this in mind we have obtained $\chi \sim 10^{-3}$, that leads to
nanotubes with around $10^{3}$ atoms each.

We will simulate the reversal process at temperature $T=300$ K, using the
scaling technique described above. MC simulations were carried out using
Metropolis algorithm with local dynamics and single-spin flip methods.\cite%
{binder} One interesting point to be considered is the effect of scaling on
temperature at which the simulations are carried out. As explained by
Bahiana \textit{et al.}\cite{PRBhilos} to keep thermal activation process
invariant under the scaling transformation, the energy barriers must also be
invariant, therefore, temperature should scale as the volume, that is, $%
T^{\prime }=\chi ^{3\eta }T$.

We perform numerical simulations for tubes characterized by $\beta $ ranging
from $0.17$ to $0.83$, starting with the saturated magnetization along the
cylindrical axis, the $\mathbf{\hat{z}}$ axis, with the external field $%
\boldsymbol{H}_{a}$ applied in the $-\mathbf{\hat{z}}$ direction. Since the
nucleation of a domain wall is more likely to occur at the ends, and we wish
to follow the propagation of a single wall,\cite{wires3} the field has not
been applied to the last $12$ nm of one end, following a pinning procedure
used in experiments with microwires. For every $\beta $ we tried $\left\vert
\boldsymbol{H}_{a}\right\vert =1.3$, $1.5$ and $1.7$ kOe, without any
significant difference in the results. At least five seeds were used for
each set of $\boldsymbol{H}_{a}$ and $\beta $.\bigskip

\textbf{Continuum Model}. Results of our simulations and previous results in
wires \cite{wires1,wires2,wires4,NielschIEEE} show three main idealized
types of MR, where $\mathbf{M}$ changes from one of its two energy minima ($%
\mathbf{M}=M_{0}\mathbf{\hat{z}}$, with energy $E^{F}$)$\ $to the other ($%
\mathbf{M}=-M_{0}\mathbf{\hat{z}}$, with energy $E^{F}$)$\ $by a path such
that the energy barrier is the difference between the energy maximum ($%
E_{\max }$) and the energy minimum. These mechanisms are illustrated in Fig.
1 and correspond to: \textit{Coherent Rotation}, $C$, where all the spins
(local magnetic moments) rotate simultaneously; \textit{Vortex Wall}, $V$,
where spins rotate progressively via propagation of a vortex domain wall;
and \textit{Transverse Wall}, $T$, where spins rotate progressively via
propagation of a transverse domain wall.

\begin{figure}[h]
\includegraphics[width=7cm,height=4cm]{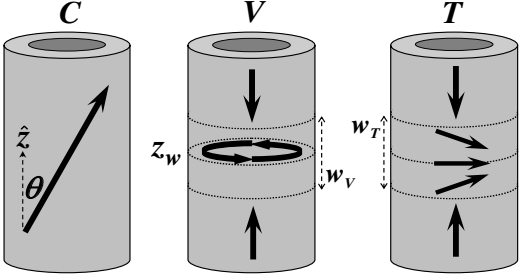}
\caption{Magnetic switching modes in nanotubes.}
\end{figure}

To determine the magnetization reversal type for a given geometry, we
calculate the energy for modes $C$, $T$ and $V$ finding the corresponding
energy barriers. We adopt a simplified description of the system in which
the discrete distribution of magnetic moments is replaced by a continuous
one, defined by a function $\mathbf{M}(\mathbf{r})$ such that $\mathbf{M}(%
\mathbf{r})\delta v$ gives the total magnetic moment within the element of
volume $\delta v$ centered at $\mathbf{r}$. The total energy for each mode
is given by the sum of exchange and dipolar contributions, which now are
taken from the well known continuum theory of ferromagnetism.\cite{Aharoni1}
The exchange term is given by $E_{ex}=A\int \sum (\boldsymbol{\nabla }%
m_{i})^{2}dv$, with $m_{i}=M_{i}/M_{0}$ ($i=x,y,z$) the cartesian components
of the magnetization\ normalized to the saturation value $M_{0}$, and $A$
the stiffness constant.\ The dipolar contribution is $E_{d}=(\mu _{0}/2)\int
\mathbf{M}\cdot \boldsymbol{\nabla }Udv$, with\ $U$ the magnetostatic
potential.\cite{Aharoni1}\ We will proceed to describe the magnetization $%
\mathbf{m}^{k}$ and evaluate the total energy $E^{k}$ for each of the three
modes defined above, with $k=C,T$ or $V$.

\noindent \textbf{Coherent Rotation (}$\mathit{C}$\textbf{). }For a
uniformly magnetized tube with $m_{z}^{C}=\cos \theta $, the exchange energy
is zero, and the total energy corresponds to the dipolar one given by%
\begin{equation*}
E^{C}(\theta )=\pi \mu _{0}M_{0}^{2}HR^{2}(1-\beta ^{2})[\sin ^{2}\theta
+(3\cos ^{2}\theta -1)N_{z}]/4\,,
\end{equation*}%
\noindent where the demagnetizing factor $N_{z}$ is given by\cite{Escrigt}%
\begin{equation*}
N_{z}=\frac{2/H}{1-\beta ^{2}}\int_{0}^{\infty }\frac{1-e^{-qH}}{q^{2}}%
[J_{1}(qR)-\beta J_{1}(\beta qR)]^{2}dq\,.
\end{equation*}%
\noindent $J_{p}$ represents a Bessel function of first kind and order $p$.
Thus for long tubes,\cite{Escrigt} $E^{C}(\theta )$ has minima at $\theta =0$
and $\theta =\pi $. That is $E^{C}(0)=E^{C}(\pi )\equiv $\ $E^{F}$
corresponds to the energy of a single domain ferromagnetic configuration
along the axis of the tube.

\noindent \textbf{Vortex domain wall (}$\mathit{V}$)\textbf{.} In this mode
the magnetization can be written as
\begin{equation*}
\mathbf{m}^{V}(z)=\left\{
\begin{array}{c}
\mathbf{\hat{z}},\ 0\leq z\leq z_{w}-w/2 \\
m_{\phi }(z)\boldsymbol{\hat{\phi}}+m_{z}(z)\mathbf{\hat{z}},\ z_{w}-w/2\leq
z\leq z_{w}+w/2 \\
-\mathbf{\hat{z}},\ z_{w}+w/2\leq z\leq H\,.%
\end{array}%
\right.
\end{equation*}%
\noindent In this expression $z=z_{w}$\ denotes the position of the center
of the domain wall - of size $w$ - that is the height at which the
magnetization lies totally in the $xy$ plane describing a perfect vortex. We
model the magnetization inside the wall with the functional $m_{z}(z)=\cos
\Theta (z)$,\ with $\ \Theta (z)=\pi \left( (z-z_{w})/w+1/2\right) $, which
describes correctly the instantaneous shape of the vortex wall in our
numerical simulations. Using $m_{x}=-m_{\phi }\left( z\right) \sin \phi $
and $m_{y}=m_{\phi }\left( z\right) \cos \phi $,\ the exchange energy
results\
\begin{equation*}
E_{ex}^{V}=\pi Aw\ln (1/\beta )+\pi ^{3}AR^{2}(1-\beta ^{2})/w\,,
\end{equation*}%
\noindent which is independent of the position of the wall, $z_{w}$. The
dipolar contribution can be written as%
\begin{equation*}
E_{d}^{V}=\pi \mu _{0}M_{0}^{2}R^{2}\int\limits_{0}^{\infty }\frac{dq}{q^{2}}%
[J_{1}(qR)-\beta J_{1}(\beta qR)]^{2}(f_{s}+f_{v})\,.
\end{equation*}%
\noindent In the above equation, $f_{s}=f_{s}(w,z_{w})$ and $%
f_{v}=f_{v}(w,z_{w})$\ are related to the surface and volumetric dipolar
energies, respectively, and are given by%
\begin{gather*}
f_{s}\equiv 1+e^{-qH}-\frac{e^{-q(H-z_{w})}+e^{-qz_{w}}}{1+q^{2}w^{2/}\pi
^{2}}\cosh [\frac{qw}{2}]\,, \\
f_{v}\equiv \frac{\frac{qw}{2}-(e^{-q(H-z_{w})}+e^{-qz_{w}})\cosh [\frac{qw}{%
2}]}{1+q^{2}w^{2/}\pi ^{2}}+\frac{1+e^{-qw}}{(1+q^{2}w^{2/}\pi ^{2})^{2}}.
\end{gather*}

\noindent \textbf{Transverse domain wall (}$\mathit{T}$\textbf{)}. Let us
orientate the $\mathbf{\hat{x}}$ axis along the transverse direction. Then,
the magnetization inside the wall (see Fig. 1) can be written as,
\begin{equation*}
\mathbf{m}^{T}(z)=m_{x}(z)\mathbf{\hat{x}}+m_{z}(z)\mathbf{\hat{z}},\
z_{w}-w/2\leq z\leq z_{w}+w/2\,,
\end{equation*}%
\noindent using the same geometrical notation as in the vortex wall, with $w$
the width of the transverse wall. We model the axial component of the
magnetization in this domain wall using $m_{z}(z)=\cos \Theta (z)$. The
exchange energy results%
\begin{equation*}
E_{ex}^{T}=\pi ^{3}AR^{2}(1-\beta ^{2})/w\,.
\end{equation*}%
\noindent The dipolar contribution gives%
\begin{equation*}
E_{d}^{T}=\pi \mu _{0}M_{0}^{2}R^{2}\int\limits_{0}^{\infty }\frac{dq}{q^{2}}%
[J_{1}(qR)-\beta J_{1}(\beta qR)]^{2}(g_{s}+g_{v})\,,
\end{equation*}%
\noindent with\
\begin{equation*}
g_{s}=\frac{q^{2}w^{2}/2\pi ^{2}}{1+q^{2}w^{2/}\pi ^{2}}\left( \frac{qw}{2}+%
\frac{1+e^{-qw}}{1+q^{2}w^{2/}\pi ^{2}}\right) +f_{s}
\end{equation*}%
and $g_{v}(w,z_{w})=f_{v}(w,z_{w})$.\ The dipolar volumetric contribution of
the $V$ and $T$ walls have the same functional form, although not
necessarily the same value due to the different wall widths of each mode,
both denoted by $w$.

With above expressions for the energies of the different MR processes it is
possible to obtain the reversal modes as a function of the geometry of the
tubes. We start by calculating the energy maximum for each path, which for
the $C$ mode is determined by $\theta =\pi /2$,\ whereas in the $V$ and $T$
modes occurs when the wall is in the middle of the tube, that is for $%
z_{w}=H/2$. The energy barrier $\Delta E^{k}$ of each mode can be calculated
as $\Delta E^{k}=E_{\max }^{k}-E^{F}$,\ with $E_{\max }^{k}$\ the energy of
the $k$-mode evaluated in their respective maximum. In cases $V$\ and $T$\
we have minimized the energy with regard to the wall width $w$, and then we
search for the reversal mode which costs less energy.
\begin{table}[h]
\caption{Results of simulations for different values of $\protect\beta$.}%
\begin{tabular}{|c|c|c|c|c|c|}
\hline
$\beta $ & $0.17$ & $0.33$ & $0.5$ & $0.67$ & $0.83$ \\ \hline
$Mode$ & $T$ & $T(V)$ & $V$ & $V$ & $V$ \\ \hline
\end{tabular}%
\end{table}

Table I shows the modes observed in our simulations for different $\beta $
values. While a $T$ mode is observed for $\beta \leq 0.33$, $V$ appeared
always for $\beta \geq 0.5$. For $\beta =0.33$ three seeds (of a total of
fifteen) leads to a mix behavior:\ The reversal process start as $T$ turning
to $V$ later on. The nucleation and propagation of the wall is monitored by
the value of $\bar{m}_{i}(z)\equiv \bar{M}_{i}(z)/M_{0}$, average value of
the components of the magnetic moment at a height $z$, relative to
saturation value. The position of the wall is determined by the maximum of $%
(1-|\bar{m}_{z}|)$ and is shown in Fig. 2.
\begin{figure}[h]
\begin{center}
\includegraphics[width=7cm,height=7cm]{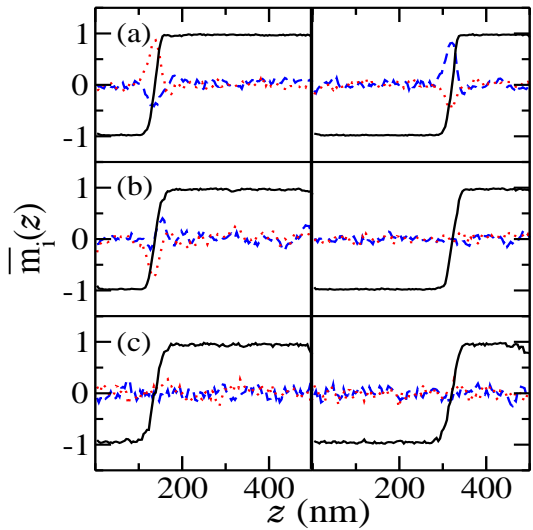}
\end{center}
\caption{(Color online). Snapshots of the reversal on tubes defined by $%
\protect\beta =0.17$ (a), $0.33$ (b) and $0.5$ (c) at two different stages
of the process. The abscissa represents the axial coordinate, $z$, along the
tube and the ordinates gives the average components of the magnetization. $%
\bar{m}_{x}$: dashed line (blue), $\bar{m}_{y}$: dotted line (red), and $%
\bar{m}_{z}$: solid line (black).}
\end{figure}
In this figure we present snapshots at two different stages of the reversal
process showing the propagation of the wall along the tube for $\beta =0.17$
(a), $0.33$ (b) and $0.5$ (c). The solid line represents the average axial
component of the magnetization ($\bar{m}_{z}$) while the other two
(in-plane) components are given by the dotted and dashed lines. When $m_{x}$
and $m_{y}$ both average to zero we face the $V$ mechanism. When one or both
of these components are non-zero, it is $T$ that is observed. Case (a) shows
a clear $T$ behavior with a helicoidal rotation of the magnetization along
the tube. Case (b) shows a $T$ mode at stage $t_{1}$, which is lost at $%
t_{2} $ where $V$ is the only mode present. The last case, (c), illustrates
a $V$ mode along the tube.
\begin{figure}[h]
\begin{center}
\includegraphics[width=7cm,height=5.5cm]{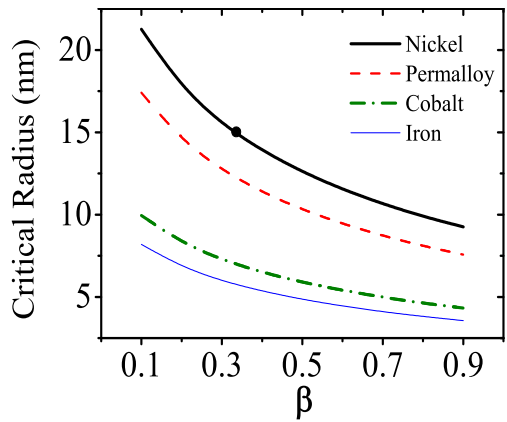}
\end{center}
\caption{(Color online). Critical radius as a function of $\protect\beta $
for different magnetic materials.}
\end{figure}

Now we turn our attention to the results provided by our theoretical model.
Upon obtaining the lowest energy barriers among the three modes we found
that $C$ is present only in very short tubes, namely, when $H\approx w$ or
less. Since we are mainly interested in long tubes, as those used in
experiments and applications mentioned above, we will focus our attention on
$T$ and $V$ only. Because it is the energy of the wall which determines the
reversal mode, if $H>w$ the height of the tube has no effect on the
reversion process. In this case, our results show that the reversal mode
depends on the internal and external radius, the stiffness constant $A$ and
the saturated magnetization $M_{0}$. For each $\beta $ there exists a
critical radius, $R_{c}(\beta )$, at which both energy barriers, $T$ and $V$%
, are equal. For $R<R_{c}(\beta )$ the tube reverses its magnetization
creating a transverse wall, while for $R>R_{c}(\beta )$ a vortex wall
appears. By equating both wall energies it is possible to obtain $%
R_{c}(\beta )$, which is illustrated in Fig. 3 for Nickel (thick solid
line), Permalloy (dashed line), Cobalt (dash-dotted line ) and Iron (thin
solid line). Parameters for Permalloy, Cobalt and Iron have been taken from
ref. 16, and for Nickel are the ones mentioned before. For $R=15$ nm in
Nickel we observe that for $\beta <0.33$ a $T$ mode occurs, a mixed reversal
behavior is observed for $\beta =0.33$ (denoted by a dot on the Ni line) and
a $V$ mode appears for $\beta $ $>0.33$. These results perfectly agree with
our numerical simulations (see Fig. 2 and Table I).

\begin{figure}[h]
\begin{center}
\includegraphics[width=8cm,height=4.5cm]{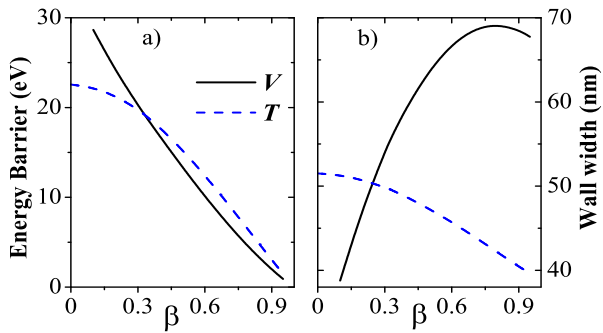}
\end{center}
\caption{(Color online). Solid lines (black) illustrate (a) the energy
barriers and (b) wall width for the $V$ mode and dashed lines (blue)
correspond to the $T$ mode.}
\end{figure}

Fig. 4 illustrates our analytical results for the Ni tubes used in our
simulations. Figure 4(a) depicts the energy barriers for the $V$ and $T$
modes as a function of $\beta $. We observed that at $\beta \approx 0.33$
the reversal mode changes from $T$ to $V$, in accordance to our previous
numerical simulations. Figure 4(b) illustrates the corresponding wall widths
for modes $T$ and $V$, which are of the order of the nanometers. Then, for
very short tubes ($H\approx w$) it is not possible to accommodate a
transverse or vortex wall, giving rise to a coherent mode of reversal.

In conclusion, there are two basic reversion processes in long ferromagnetic
nanotubes depending on the radii of the tube. The transverse mechanism is
present in tubes with $R<R_{c}(\beta )$ while for tubes with $R>R_{c}(\beta
) $ it is always the vortex mode that controls the magnetization reversal.
For values of $R$ and $\beta $ close to the boundary between the two phases,
instabilities arise and changes between $T$ and $V$ modes can occur during
the process of reversion. Because nanotubes with radius smaller than $20$ nm
are difficult to fabricate at present,\cite{tubes1,tubes2,Daub} we can
conclude that the vortex reversal mode is the most observed.

This work has been partially supported by FONDECYT under Grants Nos.1050013
and 1060317 and Millennium Science Nucleus "Condensed Matter Physics"
P02-054F of Chile. CONICYT Ph.D. Program Fellowships and MECESUP USA0108
project are also acknowledged.


\begin{thebibliography}{99}
\bibitem{tubes1} K. Nielsch, F. J. Casta\~{n}o, C. A. Ross and R. Krishnan,
J. Appl. Phys. \textbf{98}, 034318 (2005).

\bibitem{tubes2} Kornelius Nielsch, Fernando J. Casta\~{n}o, Sven Matthias,
Woo Lee, Caroline A. Ross, Adv. Eng. Mat. \textbf{7}, 217-221 (2005).

\bibitem{Daub} M. Daub, M. Knez, U. G\"{o}sele and K. Nielsch, submitted to
J. Appl. Phys. (2006).

\bibitem{tubes3} Z. K. Wang \textit{et al}., Phys. Rev. Lett. \textbf{94},
137208 (2005). \ \

\bibitem{Escrigt} J. Escrig, P. Landeros, D. Altbir, E. E. Vogel, and P.
Vargas, J. Magn. Magn. Mater. \textbf{308}, 233--237 (2007).

\bibitem{wires3} R. Varga, K. L. Garcia, M. V\'{a}zquez and P. Vojtanik,\
Phys. Rev. Lett. \textbf{94}, 017201 (2005).

\bibitem{wires1} D. Hinzke, U. Nowak, J. Magn. Magn. Mater. \textbf{221}
365-372 (2000).

\bibitem{wires2} H. Forster, T. Schrefl, D. Suess, W. Scholz, V. Tsiantos,
R. Dittrich, and J. Fidler, J. Appl. Phys. \textbf{91}, 6914 (2002).

\bibitem{wires4} R. Wieser, U. Nowak, and K. D. Usadel, Phys. Rev. B \textbf{%
69}, 064401 (2004).

\bibitem{NielschIEEE} K. Nielsch, R. Hertel, R. B. Wehrspohn, J. Barthel, J.
Kirschner, U. G\"{o}sele, S. F. Fischer and H. Kronm\"{u}ller, IEEE Trans.
Magn. \textbf{38}, 2571 (2002).

\bibitem{scaling1} J. {d'Albuquerque e Castro}, D.\ Altbir, J.\ C. Retamal,
P.\ Vargas, Phys. Rev. Lett. \textbf{88}, 237202 (2002) .

\bibitem{fastmontecarlo} P. Vargas, D. Altbir, and J. d'Albuquerque e
Castro, Phys. Rev. B \textbf{73}, 092417 (2006).

\bibitem{binder} K. Binder, D.\ Heermann, Monte Carlo Simulation in
Statistical Physics, Springer, 2002.

\bibitem{PRBhilos} M. Bahiana, F. Amaral, S. Allende and D. Altbir,\ Phys.
Rev. B (to be published).

\bibitem{Aharoni1} A. Aharoni, Introduction to the Theory of Ferromagnetism
(Clarendon Press, Oxford, 1996).

\bibitem{ohandley} R. C. O'Handley, Modern Magnetic Materials (John Wiley \&
Sons, Inc; USA, 2000).
\end{thebibliography}
\end{document}